# Large transport critical currents in dense Fe- and Ni-clad MgB$_2$ superconducting tapes

HongLi Suo, Concetta Beneduce, Marc Dhallé, Nicolas Musolino, Jean-Yves Genoud, and René Flükiger.

Département de Physique de la Matière Condensée, Université de Genève, 24 Quai Ernest-Ansermet, CH-1211 Genève 4, Switzerland

We report on the preparation of dense monofilamentary MgB$_2$/Ni and MgB$_2$/Fe tapes with high critical current densities. In annealed MgB$_2$/Ni tapes, we obtained transport critical current densities as high as $2.3 \times 10^5$ A/cm$^2$ at 4.2 K and 1.5 T, and for MgB$_2$/Fe tapes $10^4$ A/cm$^2$ at 4.2 K and 6.5 T. To the best of our knowledge, these are the highest transport j$_c$ values at 4.2 K reported for MgB$_2$ based tapes so far. An extrapolation to zero field of the MgB$_2$/Fe data gives a critical current value of ~ 1 MA/cm$^2$, corresponding to an critical current value well above 1000 A. The high j$_c$ values obtained after annealing are a consequence of sintering densification and grain reconnection. Fe does not react with MgB$_2$ and is thus an excellent sheath material candidate for tapes with self-field j$_c$ values at 4.2 K in excess of 1 MA/cm$^2$.

The recent discovery of superconductivity in $MgB_2$ at 39 K[1] has spurred a considerable volume of fundamental and applied research. High transport critical current densities ($j_c$) have been reported both in $MgB_2$ bulk samples[2, 3, 4] and in thin films[5, 6]. Due to its relatively high transition temperature, weak-link free grain boundaries[7] and low cost, this material is interesting for different applications, such as superconducting tapes. Canfield et al.[8] were the first to report on $MgB_2$ wires with inductive self-field $j_c$ values of $10^5$ A/cm$^2$ at 4.2 K. Several research groups have reported on $MgB_2$ wires or tapes using the Powder-In-Tube (PIT) method, either with[9, 10, 11] or without[12, 13] heat treatment after deformation, with transport $j_c$ values at 4.2 K and in zero external field in the range of $10^4$-$10^5$ A/cm$^2$. Different sheath materials have been used. Toughness and chemical compatibility with $MgB_2$ are reported to be crucial for fabricating long $MgB_2$ tapes with high critical currents[10, 12]. In this work, we report on high $j_c$ PIT tapes using Ni and Fe sheaths. We compare microstructure, superconducting transition and $j_c$ between the as-rolled and the recrystallized tapes, and find a significant improvement of the $j_c$ values after heat treatment.

The $MgB_2$ tapes were fabricated by sealing commercial 98% pure $MgB_2$ powder (Alfa-Aesar) into Ni or Fe tubes under an inert Ar atmosphere. The Ni tubes had an outer diameter (OD) of 12.7 mm and an inner diameter (ID) of 7 mm, while the Fe tubes had OD = 8 mm, ID = 5mm. Both ends of the tubes were sealed with lead pieces. After drawing to a diameter of 2 mm, the wires were cold-rolled to a thickness of 230 µm for $MgB_2$/Ni and 375 µm for $MgB_2$/Fe. The superconductor filling factors were 20% for the $MgB_2$/Ni tapes and 28 % for the $MgB_2$/Fe tapes. Fig. 1.a and b show scanning electron microscopy (SEM) micrographs of polished cross-sections of the as-rolled tapes, indicating a tightly packed powder with grain size of 0.5-2 µm, markedly smaller than the initial grain size due to deformation-induced grain refinement. These tapes were annealed for 0.5 h in a pure Ar atmosphere at 950 °C and 980 °C for the $MgB_2$/Fe and $MgB_2$/Ni tapes respectively. Mass losses were less than 0.2 %. Recrystallization and sintering strongly improved the intergranular connectivity, yielding denser filaments as can be seen in Fig. 1.c and d. We estimated the filament density after annealing to be close to the theoretical density of 2.62 g/cm$^3$ [14]. Sintering caused the core to shrink and the filling factors to decrease from 28 % to 26 % in the MgB$_2$/Fe. The filling factor in MgB$_2$/Ni tapes showed an even stronger decrease from 20 % to 14 %, due to the formation of a ~ 15 µm thick reaction layer consisting of Mg$_2$Ni between the sheath and the filament (Fig. 1.e). This reaction consumes some of the Mg and thus increases the porosity. No reaction was found in the MgB$_2$/Fe tape (Fig. 1.f), and X-ray measurements of its crushed core revealed ~ 5 % of MgO as the only impurity phase.

Figure 2 shows A.C. and D.C. susceptibility data for the MgB$_2$ cores before and after annealing, and for the starting powder. As-rolled and annealed tapes showed an onset of the superconducting transition at lower temperature (around 37 K) than the starting powder (38 K). This lowering of $T_c$ may be explained by deformation induced stresses, which are not completely released after annealing, as confirmed by X-ray pattern refinement. Note that *in-situ* measurements of the pressure dependence of $T_c$ have also revealed a decrease of $T_c$ upon compressive strain (0.8 - 2 K/GPa)[15, 16], being only partially recovered when the pressure is released[17]. Moreover, the deformation dramatically affects the grain connectivity as shown by the marked weak link behavior of the as-rolled tapes, indicated by a dissipative peak in $\chi''$, and confirmed by SQUID magnetometry[18]. After annealing the connectivity is recovered and the tapes present a unique and sharp superconducting transition.

Transport critical currents were measured on 45 mm long tapes pieces in a He bath at 4.2 K. The voltage contacts were 10 mm apart, and the voltage criterion used was 1 µV/cm. The magnetic field was parallel to the tapes surface and perpendicular to the current direction. The field dependence of $j_c$ in the MgB$_2$/Ni and MgB$_2$/Fe tapes, both before and after the annealing, is shown in figure 3. For comparison, the transport $j_c$ for a hot forged bulk MgB$_2$ sample[4] is also shown. Lack of thermal stabilization causes sample quenching above a given current density in most samples. The current, at which quenching occurs, was considerably higher for the $MgB_2$/Ni tapes than for the $MgB_2$/Fe tapes, presumably due to the smaller core thickness: the amount of heat generated in a bulk dissipation process is proportional to the core volume, whereas the cooling efficiency is proportional to the core surface. To fully exploit the potential of this material, it will be essential to use thinner cores or multi-filamentary tapes[13, 19].

The as-rolled $MgB_2$/Ni tape had a self-field transport $j_c$ values of $6.3 \times 10^4$ A/cm$^2$ at 4.2K. In a field of 2 T, its $j_c$ value was ~ 30 % higher than for the as-rolled Fe-sheathed tapes. As reported by Grasso et al.[12], we found an inverse correlation between the filling factor and $j_c$[20], partially explaining the difference in $j_c$ values for the as-rolled tapes. Although the sheath material and the filling factor are important parameters to obtain good $j_c$ values for as-rolled tapes, we only expect significant improvements of $j_c$ in heat treated samples.

During annealing the core material recrystallized and sintered, thus strongly improving connectivity. This lead to a significant enhancement of the transport $j_c$ values, which increased by more than a factor of 10, reaching $2.3 \times 10^5$ A/cm$^2$ in 1.5 T for $MgB_2$/Ni ($I_c$ = 300 A) and ~ $10^4$ A/cm$^2$ in 6.5 T for $MgB_2$/Fe. To the best of our knowledge, these values represent the highest transport $j_c$ so far obtained in tapes of this compound at 4.2 K. Comparison with the high-pressure sintered bulk sample, whose transport $j_c$ is essentially equal to the intragranular $j_c$[4], shows that also in our annealed tapes $j_c$ is only marginally affected by connectivity issues. Additional inductive measurements revealed that the temperature dependence of $j_c$ in the annealed tapes was very similar to that of the bulk sample[20]. The different transport $j_c$ values in the annealed $MgB_2$/Ni and $MgB_2$/Fe tapes were mainly due to chemical reasons. The reaction between Ni and $MgB_2$ disturbed the stoichiometry of the core, increased the porosity, and lowered the transport $j_c$. Extrapolating the field dependence of transport $j_c$ in Fe-sheathed tapes yielded self-field values as high as 1 MA/cm$^2$. The same high value of $j_c$ was obtained from inductive measurements performed in a vibrating sample magnetometry[20]. Such a high $j_c$ corresponds to critical current well above 1000A, which can be only measured once the thermal stability issue is resolved by using, for example, thinner filaments.

In conclusion, we prepared highly dense superconducting $MgB_2$ tapes with large transport $j_c$ using Ni and Fe sheaths. Annealing increased core density and sharpened the superconducting transition thus raising $j_c$ by more than a factor of ~ 10. Annealed $MgB_2$/Ni tapes yielded $j_c$ values up to $2.3 \times 10^5$ A/cm$^2$ at 4.2 K in a magnetic field of 1.5 T. The $j_c$ values in $MgB_2$/Fe tapes were higher, up to $10^4$ A/cm$^2$ at 4.2 K and 6.5 T, the estimated self-field value of $j_c$ at 4.2 K being close to $10^6$ A/cm$^2$. The high hardness and chemical compatibility of Fe make it a suitable sheath material for fabricating long-length $MgB_2$ tapes with a realistic potential for high-current applications.

We thank Pierre Toulemonde, Enrico Giannini, and Eric Walker for useful discussions, XiaoDong Su for helping in the PIT tapes preparation, and Robert Janiec for his help with the transport measurements.

*Figure captions*

**Figure 1:** SEM microstructures of the MgB$_2$ tapes: (a) MgB$_2$ core in the as-rolled MgB$_2$/Ni tape; (b) MgB$_2$ core in the as-rolled MgB$_2$/Fe tape; (c) MgB$_2$/Ni core after annealing; (d) MgB$_2$/Fe core after annealing; (e) reaction layer in annealed MgB$_2$/Ni tape; (f) annealed MgB$_2$/Fe tape, showing no reaction layer.

**Figure 2:** A.C. susceptibility (0.1 Oe, 8 KHz) and D.C. SQUID susceptibility (20 Oe) as a function of temperature for both as-rolled and annealed MgB$_2$ tapes. [(1): MgB$_2$/Fe annealed; (2): MgB$_2$/Ni annealed; (3) MgB$_2$/Fe as-rolled; (4): MgB$_2$/Ni as-rolled; (5) starting powder].

**Figure 3:** Field dependence of the transport j$_c$ values at T = 4.2 K in both as-rolled and annealed MgB$_2$ tapes. For comparison, the transport j$_c$ curve of a dense hot forged bulk sample is shown[4].

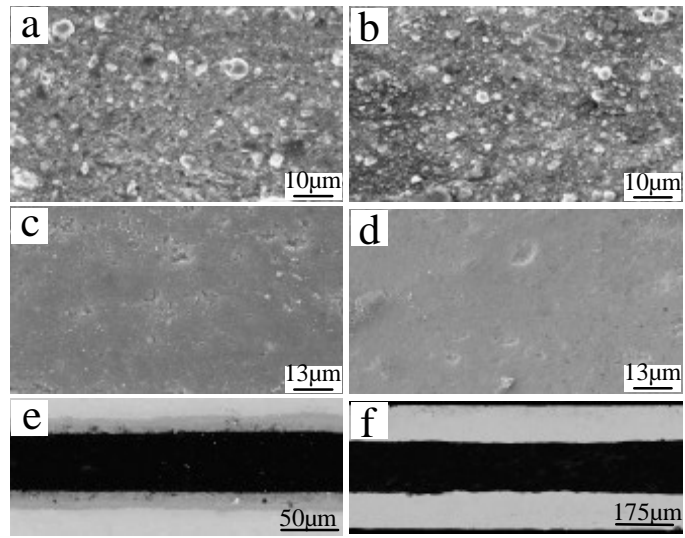

**Fig. 1**

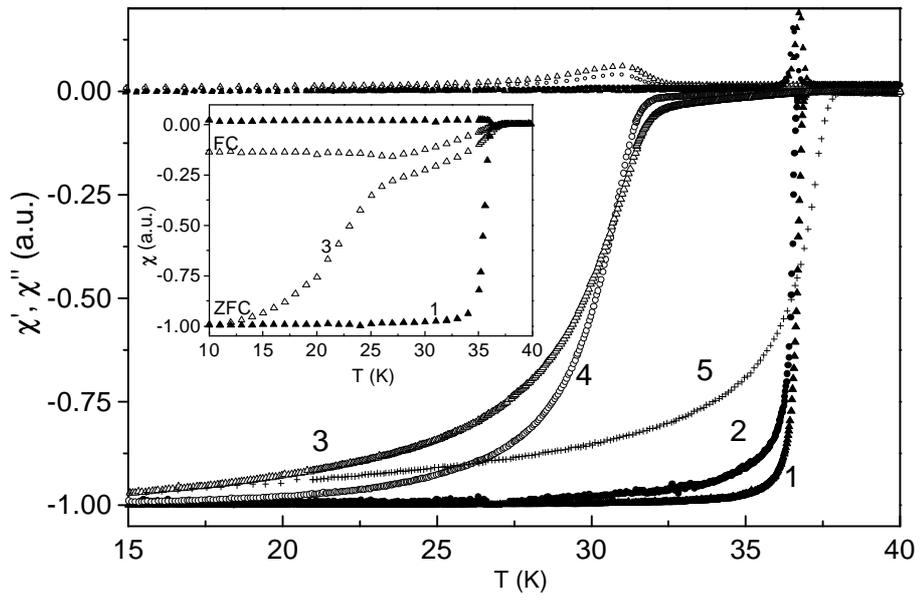

**Fig. 2**

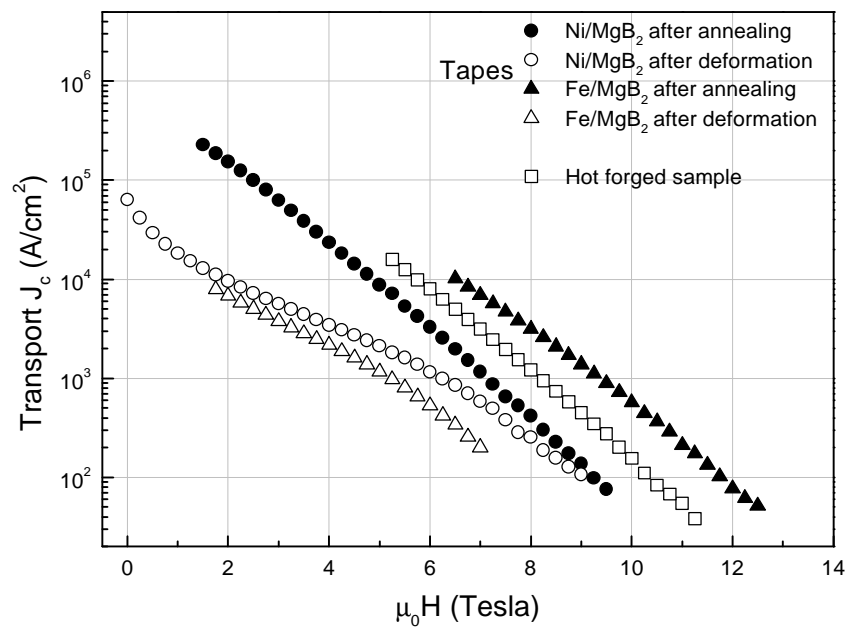

**Fig. 3**

*References*

[1] J. Nagamatsu, N. Nakagawa, T. Muranka, Y. Zenitanim, and J. Akimitsu, Nature, **410** 63 (2001).

[2] M. Kambara, N.H. Babu, E.S. Sadki, J.R. Cooper, H. Minami, D.A. Cardwell, A.M. Campbell, and I.H. Inoue, Supercond. Sci. Technol. **14**, L5 (2001).

[3] Y. Takano, H. Takeya, H. Fujii, H. Kumakura, T. Hatano, and K. Togano, Appl. Phys. Lett. **78**, 2914 (2001).

[4] M. Dhallé, P. Toulemonde, C. Beneduce, N. Musolino, M. Decroux, and R. Flükiger, accepted for publication in Physica C, cond-mat/0104395

[5] C.B. Eom, M.K. Lee, J.H. Choi, L. Belenky, X. Song, L.D. Cooley, M.T. Naus, S. Patnaik, J. Jiang, M. Rikel, A. Polyanskii, A. Gurevich, X.Y. Cai, S.D. Bu, S.E. Babcock, E.E. Hellstrom, D.C. Labalestier, N. Rogado, K.A. Regan, M.A. Hayward, T. He, J.S. Slusky, K. Inumaru, M.K. Haas, and R.J. Cava, Nature **411**, 558 (2001).

[6] M. Paranthaman, C. Cantoni, H. Y. Zhai, H. M. Christen, T. Aytug, S. Sathyamurthy, E. D. Specht, J. R. Thompson, D. H. Lowndes, H. R. Kerchner, and D. K. Christen, Appl. Phys. Lett. **78**, 3669 (2001)

[7] D.C. Larbalestier, L. D. Cooley, M. O. Rikel, A. A. Polyanskii, J. Jiang, S. Patnaik, X. Y. Cai, D.M. Feldmann, A. Gurevich, A. A. Squitieri, M. T. Naus, C. B. Eom, E. E. Hellstrom, R. J. Cava, K. A. Regan, N. Rogado, M. A. Hayward, T. He, J. S. Slusky, P. Khalifah, K. Inumaru, and M. Haas, Nature, **410,** 186 (2001).

[8] P.C. Canfield, D.K. Finnemore, S.L. Budko, J.E. Ostenson, G. Lapertot, C.E. Cunningham, and C. Petrovic, Phys. Rev. Lett. **86**, 2423 (2001).

[9] B. A. Glowacki, M. Majoros, M. Vickers, J. E. Evetts, Y. Shi, and I. Mcdougall, Supercond. Sci. Technol. **14,** 193 (2001).

[10] S. Jin, H. Mavoori, C. Bower, and R. B. van Dover, Nature **411**, 563 (2001).

[11] S. Soltanian, X.L. Wang, I. Kusevic, E. Babic, A.H. Li, H.K. Liu, E.W. Collings, and S.X. Dou, cond-mat/0105152 (unpublished).

[12] G. Grasso, A. Malagoli, C. Ferdeghini, S. Roncallo, V. Braccini, M. R. Cimberle, and A. S. Siri, cond-mat/0103563 (unpublished).

[13] H. Kumakura, A. Matsumoto, H. Fujii, and K. Togano, cond-mat/0106002 (unpublished).

[14] M.E. Jones and R.E. Marsh, J. Amer. Chem. Soc. **76**, 1434 (1954).

[15] M. Monteverde, M. Núñez-Regueiro, N. Rogado, K.A. Regan, M.A. Hayward, T. He, S.M. Loureiro, and R.J. Cava, Science **292**, 75 (2001).

[16] V.G. Tissen, M.V. Nefedova, N.N. Kolesnikov and M.P. Kulakov, cond-mat/0105475 (unpublished)

[17] S.I. Schlachter, W.H. Fietz, K. Grube, and W. Goldacker, presented at SCENET workshop on High-current Superconductors for Practical Applications, Alpbach, Austria, June 8-10, 2001.

[18] We note that the annealed $MgB_2$/Fe tape showed a small paramagnetic Meissner effect. The same behavior was also measured in Nb samples with strong pinning, P. Kostíc, A.P. Paulikas, U. Welp, V. R. Todt, C. Gu, U. Geiser, J. M. Williams, K.D. Carlson, and R.A. Klemm, Phys. Rev. B **53**, 791 (1996)

[19] C.F. Liu, S.J. Du, G. Yan, Y. Feng, X. Wu, J.R. Wang, X.H. Liu, P.X. Zhang, X.Z. Lu, L. Zhou, L.Z. Cao, K.Q. Ruan, C.Y. Wang, X.G. Li, G.E. Zhou, and Y.H. Zhang, cond-mat/0106061 (unpublished).

[20] C. Beneduce, H.L. Suo, M. Dhallé, N. Musolino, and R. Flükiger, to be published.